\documentclass[%
 reprint,
 amsmath,amssymb,
prl,
]{revtex4-2}

\usepackage{lipsum}
\usepackage{graphicx}
\usepackage{bm}
\usepackage{color}
\usepackage{times}
\usepackage{amsmath}
\usepackage{float}
\usepackage{hyperref}
\usepackage{ulem} 
\usepackage[dvipsnames]{xcolor}

\begin{document}


\title{Non-differentiable phononic dispersion curves
}
\title{Zero-frequency Kohn anomaly in phononic metamaterials}

\title{Observation of dispersion anomalies by design}

\author{Mahmoud M. Samak}
\author{Osama R. Bilal}
\email{osama.bilal@uconn.edu}
 \affiliation{School of Mechanical, Aerospace, and Manufacturing Engineering, University of Connecticut, Storrs , CT , 06269, USA.}

\date{\today}

\begin{abstract}
Band structures encode electronic, optical, and acoustic properties of matter and can serve as an essential tool in material discovery and design. Dispersion anomalies- sharp, non-standard features in the frequency-wavenumber relation- have been historically correlated with phonon-electron coupling or long-range interaction.  Through a combination of experimental, numerical, and analytical methods, we show how magnetic couplings can induce negative stiffness and sculpt dispersion relations to support  zero-frequency phonon anomalies at arbitrary, non-zero wavenumbers. Our approach enables the realization of complete wavenumber band gaps without time-modulation, electron-phonon coupling, or long-range interactions. We identify the conditions under which non-differentiable zero-frequency phonons exist away from the high-symmetry points. Our framework generalizes across monoatomic and diatomic lattices, locally resonant metamaterials, non-local systems, as well as higher dimensional crystals.  In addition, we report the first- passive or active- experimental observation of wavenumber band gaps in higher dimensions. Our work establishes a new paradigm in dispersion engineering and provides means for understanding wave-matter interaction in both the frequency and wavenumber domains.
\end{abstract}

\maketitle
Dispersion relations, or the way frequency depends on wave vectors, dictate wave behavior. The understanding and engineering of dispersion remains a major challenge in the electronic, optical, acoustic, and elastic domains. While smooth dispersion curves are expected in most cases, dispersion anomalies can arise in
Weyl Semimetal \cite{aynajian2008energy}, unconventional superconductors \cite{weber1987electron, reznik2006electron}, carbon nanotubes \cite{PhysRevB.75.035427}, graphene \cite{tse2008chirality,huang2009phonon}, graphite \cite{ferrari2007raman}, 1D conductors \cite{renker1973observation}, or electronic topological insulators \cite{zhu2012electron}. Dating back to 1959, Nobel Laureate Walter Kohn predicted anomalous softening (i. e., a sharp kink) of phonon dispersion curves at certain wave vectors, usually due to the interaction between lattice vibrations (phonons) and the conduction electrons in metals \cite{kohn1959image, piscanec2004kohn}. At a critical transition temperature, phonon energy can drop to zero causing a discontinuity in the derivative of the dispersion branch (i.e., group velocity) \cite {zhu2015classification, setty2024anharmonic, politano2015emergence, piscanec2007optical}. A few years earlier, Leon Brillouin drew an unconventional dispersion diagram with the dispersion curve arching towards zero frequency at an arbitrary value of the wavenumber, not only at the edges of the Brillouin zone (the high-symmetry points at wavemnubers equal to 0 or $\pi$). Brillouin speculated that such dispersion anomaly could be achieved by considering larger distances of interaction, beyond the nearest neighbor \cite{brillouin1946wave}. Unfortunately, to date,  no natural or rationally designed artificial materials show such dispersion anomalies under ambient conditions. Here, we unveil a new family of passive architected materials with unconventional phonon dispersion. In particular, we engineer the phonon dispersion curves to start from, end at, or land at zero frequency for any arbitrary wavenumber, by design.

\begin{figure*} 
\includegraphics[width = 0.8\textwidth]{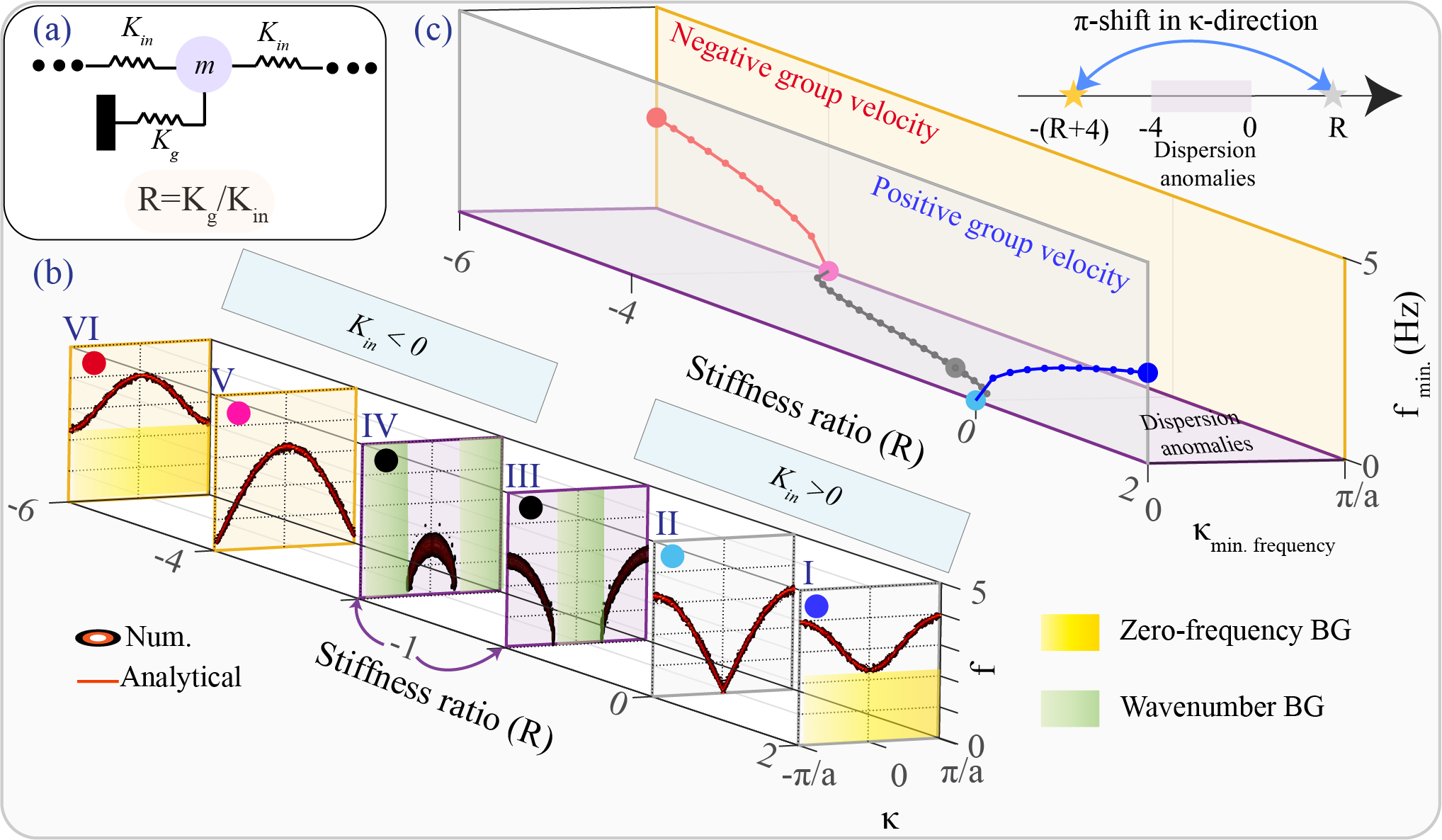}
\caption{\label{fig:concept}\textbf{Monoatomic spring-mass system.} (a) Schematic of unit cell.
(b) The evolution of dispersion curves as a function of stiffness ratio ($R=K_g/K_{in}$) calculated analytical and verified numerically as 2D-FFT overlay.
(c) The evolution of the lowest frequency in the dispersion branch and its corresponding wave number projected on group velocity planes as a function of stiffness ratios. Each filled circle represent one of the dispersion curves in (b). Inset in (c) represent dispersion dualities along the stiffness ratio axis with a phase shift of $\pi$.}
\end{figure*}

To achieve such a feat, we rely on metamaterials, which are material systems with a basic pattern that repeats in space, time, or both. Since their inception in 2000 \cite{liu2000locally}, locally resonant metamaterials broke the mass-density law by achieving negative effective dynamic stiffness in a specific frequency range \cite{zhou2012elastic, oh2016elastic, yang2021ultrawide, vo2022reinvestigation}, resulting in frequency band gaps. In addition to the band gaps in the frequency domain, few metamaterial designs can open wavenumber band gaps. Such gaps can materialize through excessive damping (only in 1D lattices) \cite{hussein2010band}, time modulation \cite{wang2018observation, chong2024modulation} and Willis-lattices \cite{ba2024onset}. Excessive non-local interactions (i.e., beyond the nearest neighbor) can soften phonon bands to a very low frequency but not to zero \cite{chen2021roton}. An unexplored avenue to engineer phonon dispersion is negative stiffness. Negative physical stiffness (rather than effective negative stiffness) can be obtained through shells \cite{tan2020mechanical}, magnetic coupling \cite{tan2019design}, Origami \cite{ma2021theoretical}, granular interaction \cite{fu2019programmable}, porous media \cite{florijn2016programmable}, or elastic buckling \cite{wang2004extreme,ha2019cubic}. Mechanical metamaterials with negative physical stiffness have been utilized for actuation, energy dissipation, or absorption \cite{haghpanah2017elastic, tang2020leveraging, tan2024negative}. However, such negative physical stiffness is not frequency dependent and can lead to instability \cite{li2020negative, ginosar2011metastability, nicolaou2012mechanical}, therefore it is not commonly used in dispersion engineering. In this paper, we formulate a general theory, that we verify experimentally, regarding the utilization of negative physical stiffness along with local instabilities to design for phonon dispersion anomalies. 

Our design methodology combines negative physical inter-stiffness and ground-stiffness to enable the realization of zero-frequency phonons at non-zero wavenumbers.  We start with a simple monoatomic lattice with ground stiffness to derive the necessary conditions for the targeted dispersion anomalies. Afterwards, we utilize macroscopic magnetic lattices to experimentally validate our hypothesis. Furthermore, we generalize our framework to more advanced lattices such as diatomic phononic crystal, locally resonant metamaterials, monoatomic lattice with non-local interactions, in addition to higher dimensional lattices. Finally, we validate our theory through the experimental realization of two-dimensional lattices with dispersion anomalies.

\begin{figure*} 
\includegraphics[width = 0.8\textwidth]{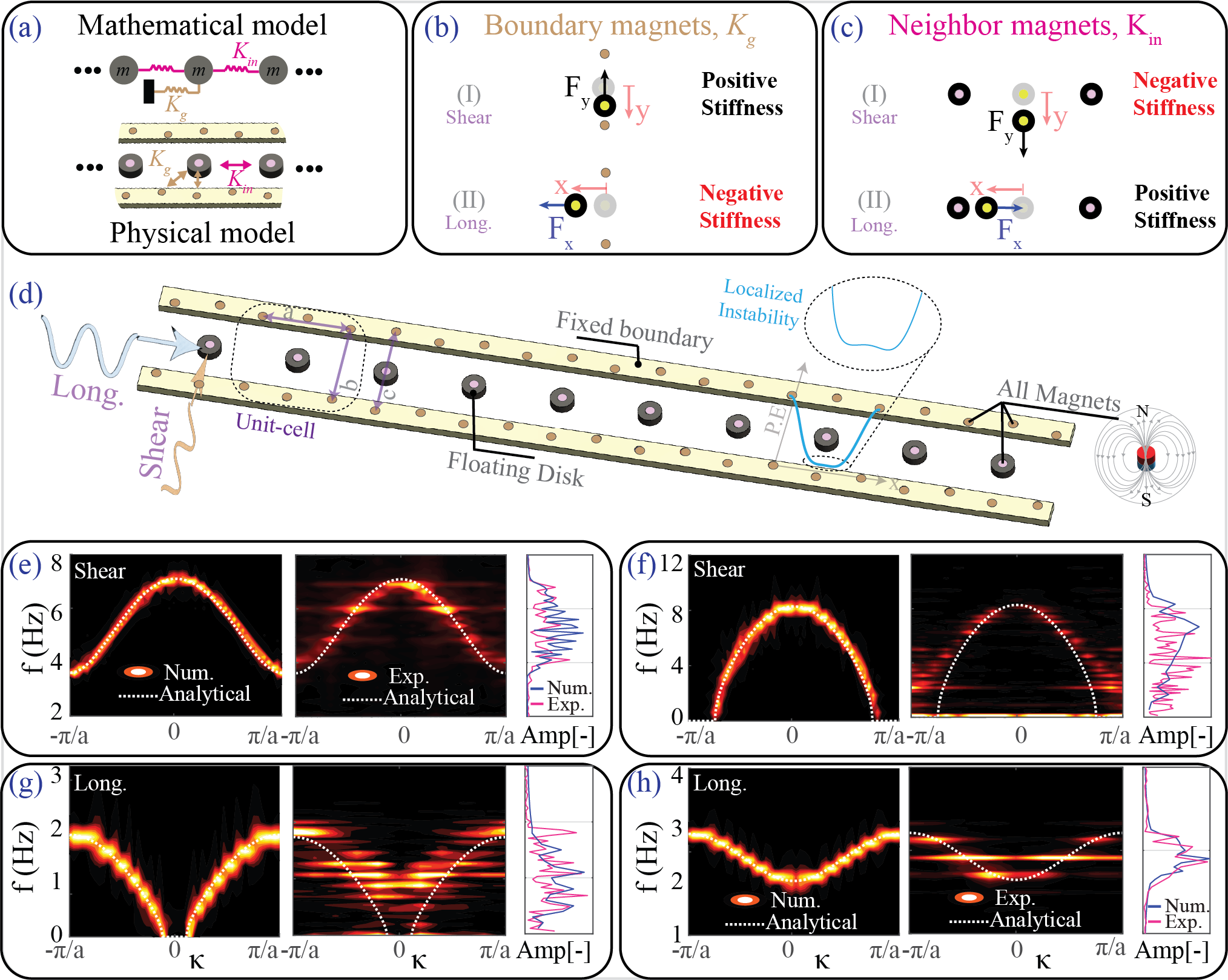}
\caption{\label{fig:1D magnetic }\textbf{Experimentally observed anomalies in magnetic lattices (1D).} 
(a) Schematic of the physical model and the corresponding mathematical model. (b) Boundary magnets' arrangement to obtain (I) positive ground stiffness in shear and (II) negative ground stiffness in longitudinal direction. (c) Neighbor magnets' arrangement to obtain (I) negative inter-stiffness in shear  and (II) positive inter-stiffness in longitudinal direction. (d) Detailed schematic of the physical model showing both excitation directions (i. e.,  longitudinal and shear) and the potential energy of the disk as a function of disk location.
Analytical, numerical and experimental dispersion curves for designs with (e) R=-5.4, (f) R = -3.4, (g) R = -0.13, and (h) R = 8.5. In panels (e-f) [Left] Numerically validated dispersion curves. [Middle] Experimentally validated dispersion curves. [Right] The transmission region using 1D FFT both numerically and experimentally.}
\end{figure*}

\begin{figure*} 
\includegraphics[scale = 0.95]{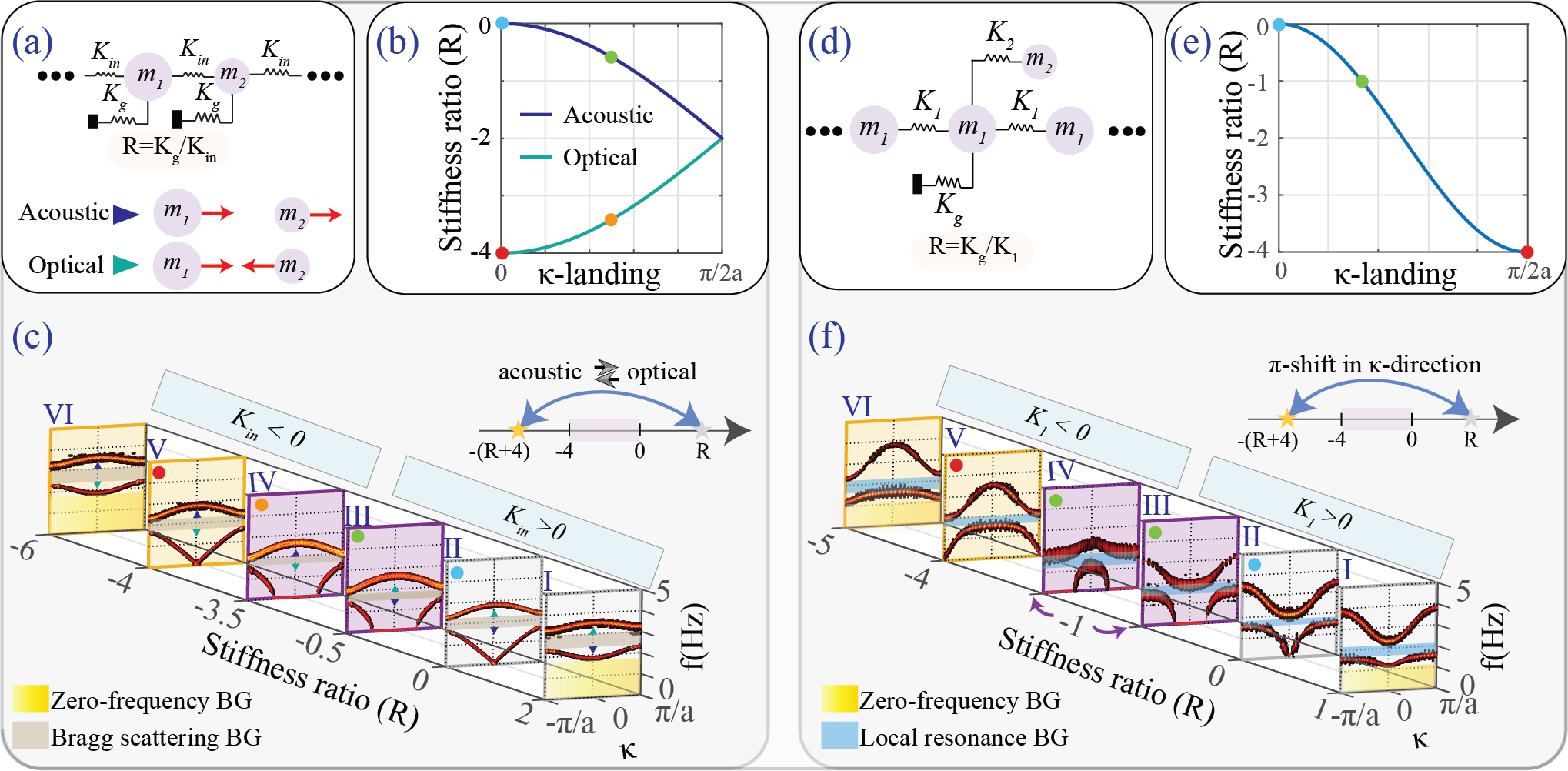}
\caption{\label{fig:diatomic}\textbf{Diatomic spring-mass system.} (a) Schematic of a phononic crystal unit cell. 
(b) The landing wavenumber as a function of the stiffness ratio, \textit{R}. (c) The evolution of dispersion curves as a function of stiffness ratio ($R=K_g/K_{in}$) calculated analytical and verified numerically as 2D-FFT overlay. Inset in (c) represent dispersion dualities along the stiffness ratio axis with a swap between acoustic and optical branches.
(d) Schematic of an acoustic metamaterials unit cell. (e) The landing wavenumber as a function of the stiffness ratio within the dispersion anomalies region. (f) The evolution of dispersion curves as a function of stiffness ratio ($R=K_g/K_{in}$) calculated analytical and verified numerically as 2D-FFT overlay. Inset in (f) represent dispersion dualities along the stiffness ratio axis with a phase shift of $\pi$.}
\end{figure*}

\section{Theory of dispersion anomalies}
 
In a conventional dispersion curve of a monoatomic lattice without ground stiffness (Fig.\ref{fig:concept}-b(II)), the dispersion branch starts from zero-frequency and zero-wavenumber reaching a maximum cut-off frequency with a positive slope (i.e., positive group velocity). Ground stiffness can be an avenue for dispersion engineering \cite{wang2018controllable}. A small amount of ground stiffness can lead to a zero-frequency band gap \cite{yang2024experimental}, while excessive values of ground stiffness can lead to complete flattening of the dispersion branches \cite{samak2024direct}. Upon the addition of ground stiffness to the lattice (Fig.\ref{fig:concept}-a), the dispersion relation can diverge into three possible scenarios:

In $scenario~1$, where both inter- and ground stiffness are positive, the dispersion curve exhibits a zero-frequency band gap and positive group velocities for all wavenumbers (Fig.\ref{fig:concept} (b)I) (the gray plane in Fig.\ref{fig:concept}(c)). 

When either the inter- or the ground stiffness is negative, but not both, this takes us into either scenario 2 or 3.  In $scenario~2$ ({$K_{in} <$ 0}), where the ratio, $R$, between inter- and ground stiffness is less than -4,  the dispersion curve exhibits a zero-frequency band gap; however, the group velocity is negative for all wavenumbers. (Fig.\ref{fig:concept}(b)V-VI) (the orange plane in {Fig.}\ref{fig:concept}(c)). In the special case of $R= -4$ the zero-frequency gap closes, making the dispersion an exact inversion of the conventional dispersion at $R=0$. It is worth noting that the dispersion curve for lattices with $R \ge 0$ from scenario 1 have identical dispersion curves to scenario 2 dispersions with $R \le -4$ in frequency range and group velocity value, yet  opposite group velocity sign. 

 In the intriguing $scenario~\#$ 3, with either inter- or ground stiffness being negative and their ratio, $R$, satisfying $0 < K_g/K_{in} < -4$, the zero-frequency band gap closes and a more interesting band gap opens, a wavenumber or momentum band gap. This momentum gap emerges for select wavenumbers inside the first Brillouin zone corresponding to imaginary frequencies (Fig.\ref{fig:concept} (b)III-IV) (the purple plane in Fig.\ref{fig:concept}(c)). Both the location and width of the wavenumber band gap depend on the absolute value of the stiffness ratio, in addition to the specific sign of each stiffness (i.e., which stiffness is positive) (Fig.\ref{fig:concept} (b) III-IV). The critical wavenumber(s) corresponding to imaginary frequencies represent the intersection point(s) of the dispersion curve with the wavenumber axis. This critical wavenumber $\kappa-${landing} can be predicted precisely using our analytical model (See SI section II). 
Moreover, the wavenumber band gap can be engineered to extend from 0 to $\kappa-${landing}, by setting $K_{in}$ to a positive value (Fig.\ref{fig:concept} (b)III) or from $\kappa-${landing} to ${\pi}$/${a}$  for negative $K_{in}$  (Fig.\ref{fig:concept} (b)IV). In all plotted cases, we superimpose the numerical results on top of the analytically obtained dispersion curves in Fig.\ref{fig:concept}(b) showing a perfect match between analytical and numerical dispersion models.

\section{Experimental validation of dispersion anomalies in 1D magnetic lattices}

To validate our theory experimentally, we need to translate the mathematical model into a physical one (Fig.\ref{fig:1D magnetic }a). The challenge here is to obtain physical inter- and ground stiffness values that can vary from positive to negative and have the flexibility to be tuned to a specific ratio. To that end, we utilize free-floating magnetic disks \cite{watkins2020demultiplexing, watkins2021exploiting, norouzi2021classification, eichelberg2022metamaterials, watkins2022harnessing, samak2024evidence, stenseng2025bi} confined within a fixed magnetic boundary (Fig.\ref{fig:1D magnetic }-a). Each disk is free to oscillate on an ``Air hockey table" to minimize friction in both the x-direction (i. e., longitudinal) and the y-direction (i. e., shear). The in-plane magnetic repulsion between the boundary magnets and the free floating disks constitutes the ground stiffness, while the repulsion among the floating disks resembles the inter-stiffness. For a floating disk at equilibrium between two boundary magnets, an up- or down motion towards either of the fixed magnets results in a force opposite to the direction of the motion (i. e., positive stiffness) (Fig.\ref{fig:1D magnetic }-b (I)). However, for the same disk moving either to the left or the right, the resultant repulsion force takes place in the same direction of the motion (i. e., negative stiffness)(Fig.\ref{fig:1D magnetic }-b (II)). For a floating disk at equilibrium between two other floating disks, similar positive or negative stiffness can be obtained (Fig.\ref{fig:1D magnetic }-c) (See SI
section I).

As an experimental evidence, we consider four different designs to validate the different scenarios outlined by our theory in the previous section. The ratios between the inter- and the ground stiffness are $R=$ -5.4, -3.4, -0.13, and +8.5, respectively (Fig.\ref{fig:1D magnetic } e-h). For $R=$ -5.4, the dispersion shows a zero-frequency band gap and a negative group velocity (Fig.\ref{fig:1D magnetic } e). For $R=$ -3.4, the dispersion shows a wavenumber band gap starting at $\kappa-${landing} and ending at $\pi/a$, in addition to a negative group velocity (Fig.\ref{fig:1D magnetic } f). For $R=$ -0.13, the dispersion also shows a wavenumber band gap, although starting from 0 and ending at $\kappa-${landing}, in addition to a positive group velocity (Fig.\ref{fig:1D magnetic } g). For $R=$ +8.5, the dispersion shows a zero-frequency band gap and a positive group velocity (Fig.\ref{fig:1D magnetic } h). For each case, the transmission at the end of the finite lattice is evaluated both numerically and experimentally. In addition, the analytical dispersion is plotted against both the numerically computed and experimentally measured dispersion curves (See SI
section III).

\section{Dispersion anomalies in dispersions with more complicated structures}
To demonstrate the generality of our hypothesis, we extend our theory to more complicated dispersion curves, (1) diatomic lattices with Bragg scattering band gaps, (2) locally resonant lattices with sub-Bragg band gaps, and (3) non-local lattices with interactions beyond the nearest neighbor. 

\paragraph{\textbf{Phononic crystals}} We consider a diatomic unit cell with two masses $m_1$ and $m_2$ connected by inter stiffness $K_{in}$ and grounded by $K_g$ (Fig.\ref{fig:diatomic}-a). Following the same approach, we identify dispersion anomalies when $-4 < R < 0$ where a partial wavenumber band gap opens in either the optical or the acoustic branch [Fig.\ref{fig:diatomic}-c (III-IV)]. It is important to note that $R = -2$ is a self-dual point \cite{vitelli2020dualities}. In other words, the dispersion curves before and after $R = -2$ are identical; however, the mode shapes flip from a lower acoustic branch with all masses oscillating in-phase to a higher one (Fig.\ref{fig:diatomic}-c) (See SI section II).\\ 

\paragraph{\textbf{Acoustic metamaterials}} We consider a main mass $m_1$ and an attached local resonator $m_2$. The main masses are connected to each other by spring $K_1$, to the ground by $K_g$, and to the resonator by $K_2$ (Fig.\ref{fig:diatomic}-d).  Dispersion anomalies exist for this metamaterial when $-4 < R <0$  (Fig.\ref{fig:diatomic}-f(III-IV)). We note that for any two lattices with stiffness ratios $R$ and $-(R+4)$, the  transmission frequency range is identical, but the group velocity sign is flipped (Fig.\ref{fig:diatomic}-f) (See SI section II). 

\begin{figure} 
\includegraphics{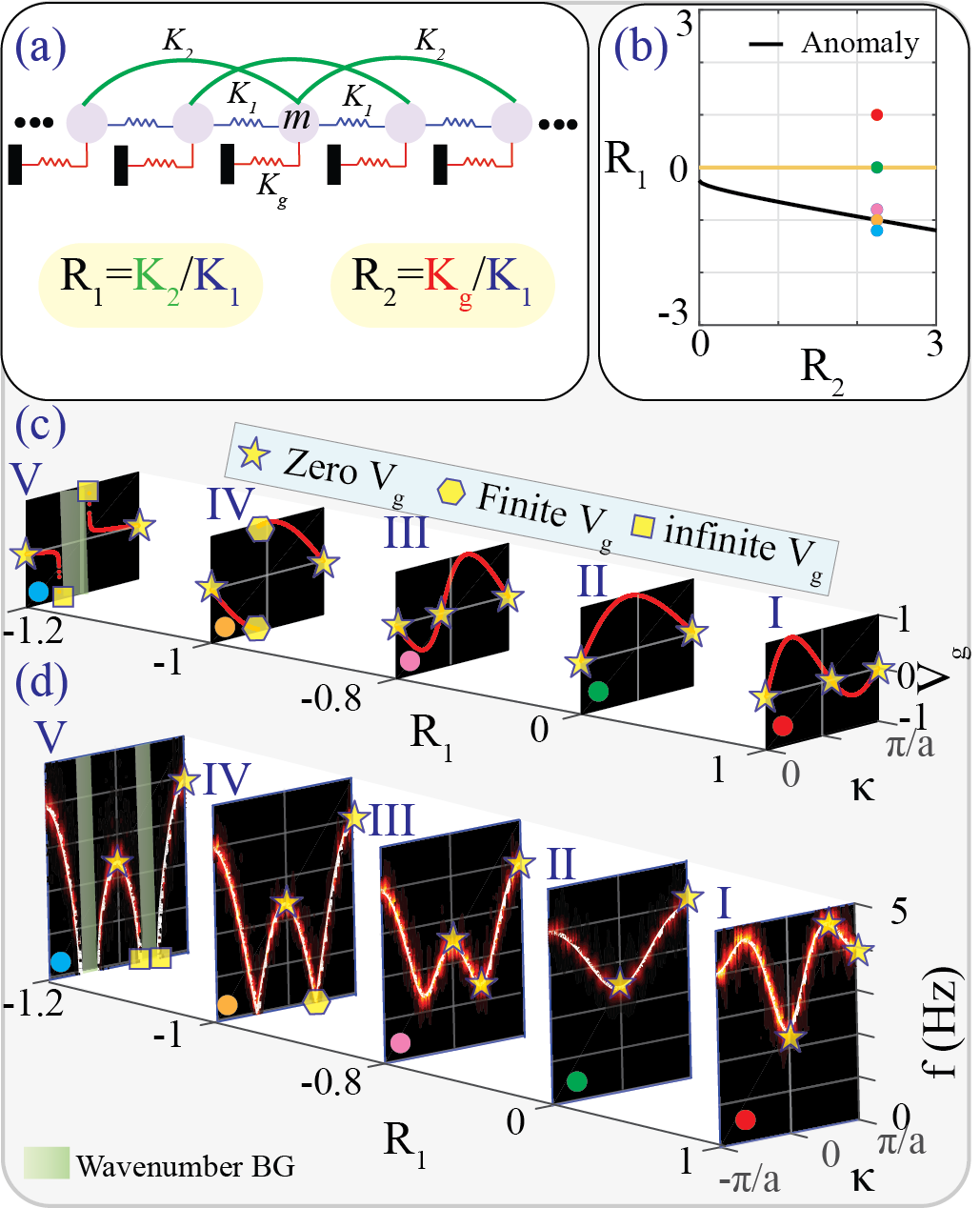 }
\caption{\label{fig:non-local}\textbf{Non-local monoatomic spring-mass system.} 
(a) Schematic of the unit cell. (b) Relation between the two stiffness ratio within the unit cell $R_1$ and $R_2$. The black line represents the perfect extreme Kohn anomaly (i. e., W-shape dispersion curve), while the orange line represents the lack of non-locality. The dots correspond to the five different cases with $R_2=2.25$ plotted in (c\&d). Evolution of (c) group velocity and (d) dispersion curves as a function $R_1$ with fixed $R_2=2.25$ calculated analytical and verified numerically as 2D-FFT overlay.}
\end{figure}

\paragraph{\textbf{Non-local metamaterials}} We consider a grounded monoatomic chain with an additional non-local stiffness $K_2$ to the second nearest neighbor (Fig.\ref{fig:non-local}-a). The dispersion relation in the presence of non-locality can be expressed in terms of two stiffness ratios $R_1={K_2}/{K_1}$ and $R_2={K_g}/{K_1}$. Without loss of generality, we set $R_2=2.25$ and vary $R_1$ (Fig.\ref{fig:non-local}-b) to reveal the intriguing dynamics of the anomalies with long-range interactions. In addition to the dispersion curves (Fig.\ref{fig:non-local}-d), we plot the corresponding group velocities for each point on the displayed dispersion curves (Fig.\ref{fig:non-local}-c). At $R_1 =0$, i. e., in the absence of long-range coupling, the dispersion exhibits zero-frequency band gap with positive group velocity (Fig.\ref{fig:non-local} d-II). When $R_1$ is positive, the conventional dispersion curve in this case becomes a Maxon-like dispersion (Fig.\ref{fig:non-local} d-I). Negative non-local coupling inverts both the Maxon-like dispersion (Fig.\ref{fig:non-local} d-III) and the sign for the group velocity (Fig.\ref{fig:non-local} c-III). At $R_1 =-1$, the dispersion branch start with a finite frequency at zero-wavenumber, decays towards zero-frequency in the middle of the first Brillouin zone, maintains zero-phonon-energy for a single wavenumber value, then rises up to a finite frequency again. This causes a discontinuity in group velocity and guarantees the formation of an extreme Kohn anomaly dispersion curve (Fig.\ref{fig:non-local} d-IV). The required stiffness ratio for this anomaly is $ R_1 \le -0.25 ~~and~~  R_2=-2[{1}/{8R_1}+2R_1+1]$ which is presented with the black line in (Fig.\ref{fig:non-local}-b). As $R_1$ decreases below -1, a wavenumber band gap opens in the middle of the first Brillouin zone (Fig.\ref{fig:non-local} d-V) (See SI section IV). 

\section{Dispersion anomalies in 2D lattices}

\begin{figure}[!h] 
\includegraphics{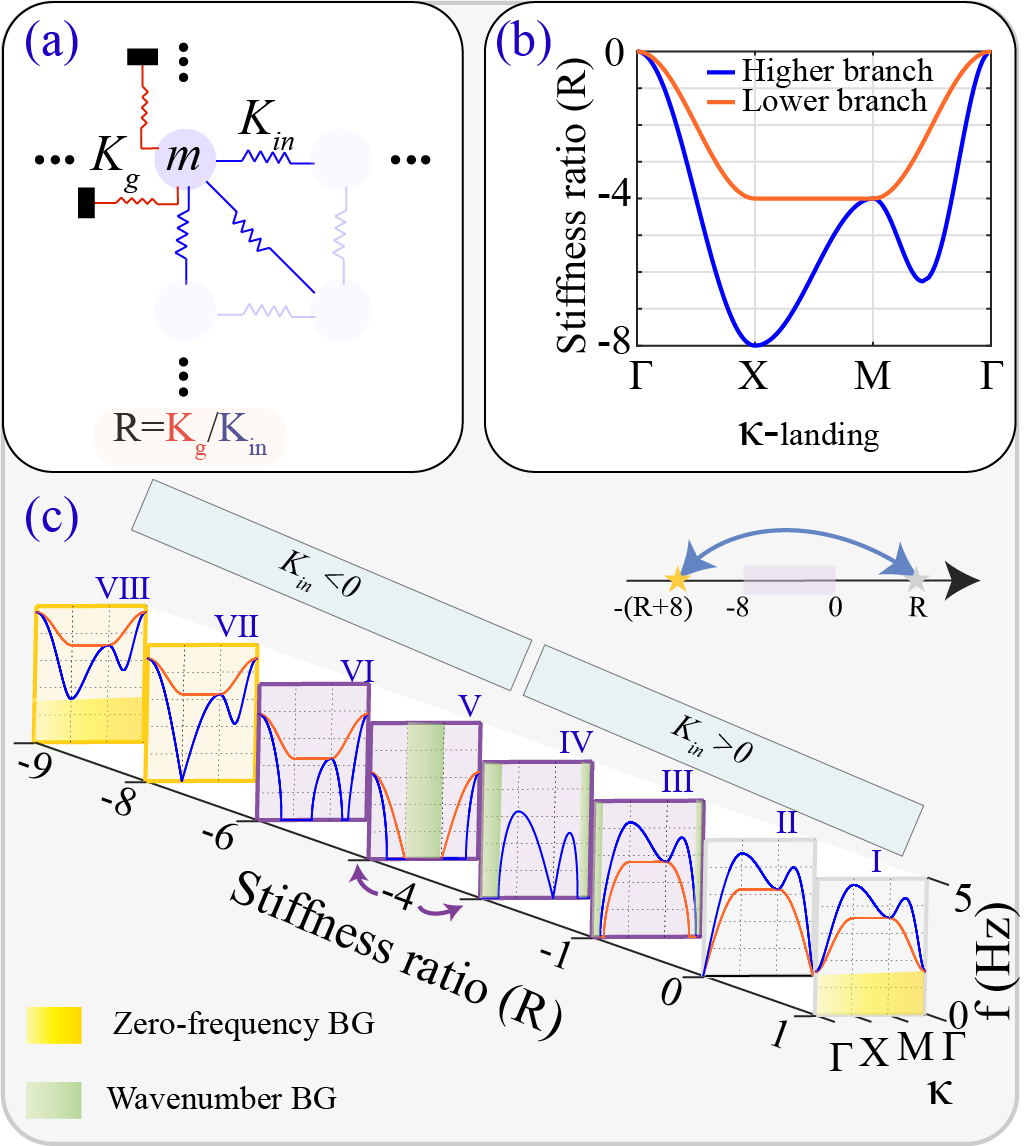}
\caption{\label{fig:2D mass spring system}\textbf{Two-dimensional spring-mass system.} (a) Schematic of the unit cell. (b) The landing wavenumber as a function of the stiffness ratio for each branch on each segment of the dispersion. (c) The evolution of dispersion curves as a function of stiffness ratio ($R=K_g/K_{in}$). Inset in (c) represent dualities in transmission frequency ranges along the stiffness ratio axis.}
\end{figure}

To show the applicability of our theory in higher dimensions, we further expand our hypothesis to 2D lattices. We consider a 2D monoatomic square lattice, where each unit cell contains a mass $m$ connected only to its nearest neighbors in all directions (i. e., horizontally, vertically, and diagonally) by an inter-stiffness $K_{in}$ and grounded both horizontally and vertically with a ground stiffness $K_g$ (Fig.\ref{fig:2D mass spring system}-a). The lattice in this case contains two dispersion curves, acoustic and optical, spanning three different segments of wavenumbers between the three high symmetry points $\Gamma, X$, and $M$. A conventional dispersion curve in two-dimensions, starts from zero-frequency and zero-wavenumber at the $\Gamma$ point and rises systematically in frequency and wavenumber from $\Gamma$ until it reaches the $X$ point at a finite frequency. After which it takes either a positive or negative slope, staying at a finite frequency value until it reaches the $M$ point. After $M$, it starts to decay until it reaches zero frequency and zero wavenumber again, at the $\Gamma$ point (Fig.\ref{fig:2D mass spring system}-c-II). By adding positive ground stiffness, both dispersion branches shift to higher frequencies, opening a zero-frequency band gap, while maintaining their shapes (Fig.\ref{fig:2D mass spring system}-c-I). To elucidate the conditions for dispersion anomalies in two-dimensional lattices, we consider the effect of $R$ (the ratio between ground- and inter stiffness) on each of the two branches along the three segments ($\Gamma - X$, $X-M$, and $M-\Gamma$). In the first segment, $\Gamma - X$, the dispersion can start from a finite wavenumber value  (Fig.\ref{fig:2D mass spring system}-c-III) or collapse to a zero-phonon energy value before the end of the segment (Fig.\ref{fig:2D mass spring system}-c-IV) when $-4 < R < 0$ for the lower branch and $-8 < R < 0$ for the higher branch. This is intriguing as it shows the possibility for opening a complete wavenumber in two-dimensional phononic dispersion for the first time. For the $X-M$ segment, an anomaly can manifest itself in the higher branch for $-8 < R < -4$ and $R = -4$ for the lower branch. Translating to a singular stiffness ratio point at which a complete wave number band gap can be opened. Lastly, the $M-\Gamma$ segment can have a zero-phonon energy away from the high symmetry points in the lower branch when $-4 < R < 0$ and in the higher branch when $-6.25 < R < 0$ (Fig.\ref{fig:2D mass spring system}-b) (See SI section V). 

\begin{figure}
\includegraphics{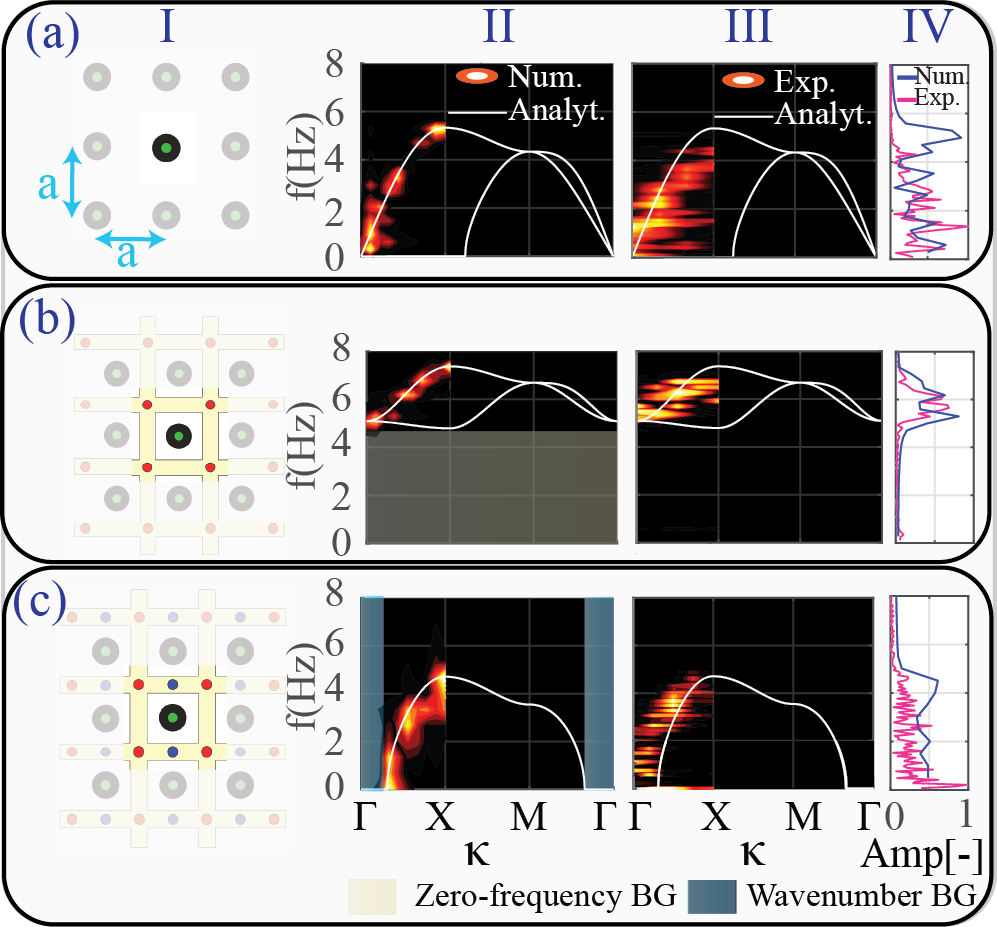}
\caption{\label{fig:2D magnetic}\textbf{Experimentally observed anomalies in magnetic lattices (2D).} Three experimentally tested design with (a) R= 0, (b) R=+4, and (c) R=-0.5. Column I in all panels includes a schematic of the tested unit cell for each design. Column II, shows the analytical dispersion curves along with the numerically computed 2D-FFT using stable-mode superposition excitation for a finite structure of 400 disks. Column III shows the analytical dispersion curves along with the experimentally obtained dispersion curves calculated from 2D-FFT for a finite structure of 99 disks and excited using a chirp signal. Column IV shows both the numerical and experimental transmission region using 1D-FFT.}
\end{figure}

\section{Experimental validation of dispersion anomalies in 2D magnetic lattices}

To experimentally validate our theory for two-dimensional lattices, we utilize the previously outlined magnetic framework with tailored stiffness. We consider three different designs to validate the three scenarios outlined by our analytical predictions in the previous section. The ratios between the ground and inter-stiffness are $R =$ 0, +4 and -0.5. For $R = 0$ (i.e., no ground stiffness), the dispersion branches start at zero, without any wavenumber or frequency band gaps (Fig.\ref{fig:2D magnetic} (a)). For $R = +4$, both dispersion branches move to finite frequencies for all wavenumbers, opening a zero-frequency band gap (Fig.\ref{fig:2D magnetic} (b)). For $R = -0.5$, the lower dispersion branch starts at a finite wavenumber with a positive group velocity, opening a wavenumber band gap (Fig.\ref{fig:2D magnetic} (c)). For each case, the transmission at the end of the finite lattice is evaluated both numerically and experimentally. In addition, the analytical dispersion is plotted against both the numerically computed and experimentally measured dispersion curves. The analytical, numerical, and experimental observations confirm our theory (See SI section VI). 
    
\section{Discussion}
The understanding and sculpting of dispersion diagrams for electrons, photons, and phonons is a rich problem with a potential impact on how we characterize and design advanced materials and devices. In this work, with a combination of analytical, numerical and experimental studies, we postulate a methodology to obtain extreme dispersion anomalies without the need for electron-phonon coupling as outlined by Kohn or long-range coupling as drawn by Brillouin. Instead, our design approach is rooted in the utilization of negative physical stiffness. We show how to obtain negative physical stiffness using magnetic particles in specific arrangements and its effect on the dispersion curves. Within our framework, we show the possibility of opening wavenumber band gaps in one-dimensional lattices without the need for active media (in contrast to time-modulated lattices or Wills lattice). We also demonstrate, for the first time, the opening of wavenumber band gaps in two-dimensional lattice, a feat that has not been reported for either passive or active media. While $\omega$-gap modes always decay in space and do not propagate, $\kappa$-gap modes exhibit exponential growth in time, leading to numerical instabilities \cite{ren2025observation}, which renders them difficult to design for and observe.  Despite these unstable modes, we outline a methodology to excite and measure such unstable wavenumber band gaps. We show the generality of our hypothesis to phononic crystals with Bragg scattering dispersion, locally resonant metamaterials, and even non-local media with long-range coupling as well as two-dimensional lattices. 

Our dispersion curves are unconventional for multiple reasons. First, the transmission region can be engineered to start from zero-frequency despite the existence of the ground stiffness. Second, the phonon energy can reach zero at $\kappa-${landing} away from the high symmetry points, even in higher dimensions (i.e, $\Gamma$, X or M) which is similar to the extreme cases of Kohn anomaly in the dispersion relation of a phonon dispersion branch in 1D metal. Third, our framework shows that a wavenumber band gaps can be opened using a new passive mechanism as opposed to the two active paradigms in the literature. Fourth, the phonon branches are wavenumber-gaped for all frequencies even in higher dimensions, something that has not been reported before for either passive or active lattice.  Our work paves the way for a completely new family of dispersion curves with anomalies in $\omega$-domain and $\kappa$-domain by providing the necessary conditions for each case and showing how to validate them in a finite structure through stable modes superposition and engineering localized bi-stability.

\section{Acknowledgment}
{This work was funded under contract W56HZV-21-2-0001 (US Army DEVCOM Ground Vehicle Systems Center (GVSC)), through the Virtual Prototyping of Autonomy Enabled Ground Systems (VIPR-GS) program and the Air force Research Labs, Materials and Manufacturing Directorate (AFRL/RXMS) contract/grant No. FA8650–21–C5711.  DISTRIBUTION STATEMENT A. Approved for public release; distribution is unlimited. OPSEC9644}

\section{Methods}
\paragraph{\textbf{Analytical model}}
The inter and ground magnetic stiffness are calculated as:
\begin{multline}
    \mathbf{K}_{in}=-\left [f_{,d} \mathbf{e}_\otimes \mathbf{e}
+ \frac{f(d)}{d}(\mathbf{I}-\mathbf{e}\otimes \mathbf{e})\right] \\
\textbf{K}_g=-\left [ \sum_{i=1}^{6}f_{,d}\textbf{e}_{i}\otimes \textbf{e}_{i}
+ \frac{f(d)}{d}(\textbf{I}-\textbf{e}_{i}\otimes \textbf{e}_{i})\right]
\end{multline}

Each matrix is a diagonal (i.e., decoupled modes) 2x2 matrix where element(1,1) represents the longitudinal stiffness and element(2,2) represents the shear stiffness.
\newline
\paragraph{\textbf{Numerical simulations}}
To validate our theory numerically, we consider a finite 1D structure ($N=50 ~ masses$). To overcome the inherent numerical instabilities and plot the band structure numerically, we excite the structure through the superposition of its stable mode shapes as the initial displacement. Then, we release the masses to vibrate freely. The initial displacement of each mass are calculated as:
\begin{equation}
    X_{i,t=0} =\sum_{j=1}^{n} U_{i,j}~~~(i=1:N) 
\end{equation}
where \textit{n} is the number of stable modes (i. e., excluding $\kappa$-gap modes)  and \textit{U} represents the eigen modes. Then we use the Verlet numerical integration method \cite{press1992numerical} to calculate the time response of each disk and apply 2D-FFT on the spatio-temporal results.
For 2D structures, we consider 400 masses (20 rows $\times$ 20 columns). We excite the structure with the initial displacements equivalent to the superposition of the stable modes on the $\Gamma-X$ branch and analyze the spatio-temporal displacement of an entire row.
\newline
\paragraph{\textbf{Experimental measurements}}
To validate our theory experimentally for 1D structures, we consider a finite structure of 20 disks. Each disk (m=0.34 g) has an embedded magnet and is surrounded by multiple boundary magnets. All magnets are 3$\times$3 and  have the same polarization and magnetic constants ($A=6.7319\times10^{-11} N/m^\gamma$, $\gamma$ = -4). We excite the first disk with a chirp signal using an electromagnet and record the motion of each disk with an eye-bird view camera. We analyze the recorded images using the digital image correlation engine DICe and perform 2D-FFT to plot the dispersion curves experimentally.

 To validate our theory experimentally in 2D, we consider a 9 x 11 square lattice (a=20 mm) of each design. Each disk has a mass $m=0.34 g$ including the embedded $3 \times 3$ magnet. We excite a disk on one of the structure edges with a chirp signal using an electromagnet and analyze the motion of all disks in the considered row. We repeat the experiment three times, first excitation from 0 Hz to 1 Hz, second from 1 Hz to 2 Hz and finally from 2 Hz to 10 Hz to cover the entire frequency domain. 
 \newline
 \paragraph{\textbf{Unit-cell design}}
In \textbf{Design 1}, we consider the shear motion in a unit-cell with a=14 mm, b=30 mm and c $\rightarrow \inf$  which provides a stiffness ratio, R=-5.4 (Fig.\ref{fig:1D magnetic }-e). In \textbf{Design 2}, we focus on the shear motion in a unit-cell with a=12 mm, b=29 mm and c $\rightarrow \inf$ to get a -3.4 stiffness ratio (Fig.\ref{fig:1D magnetic }-f). In \textbf{Design 3}, we consider the longitudinal motion of a disk in a unit-cell with a=b=30 mm and c=32.9 mm to achieve a -0.13 stiffness ratio (Fig.\ref{fig:1D magnetic }-g). In \textbf{Design 4}, we consider the longitudinal vibration in a unit-cell with a=b=30 mm and c $\rightarrow \inf$ to target a 8.5 stiffness ratio (Fig.\ref{fig:1D magnetic }-h). To avoid the global instability due to the wavenumber band gaps (i.e., in 1D structures with \textbf{Design 2} and \textbf{Design 3} and 2D lattice with R=-0.5), we introduce local instability within each unit-cell. These instabilities can be achieved by having an almost constant potential energy at the middle  of the unit-cell (Fig.\ref{fig:1D magnetic }-f) or two potential wells with very small energy barrier in-between (Figs.\ref{fig:1D magnetic }-g and \ref{fig:2D magnetic}-c). Our strategy can guarantee the global stability of the lattice and localized instability within the unit-cell.

For 2D unit cells, we consider three designs: (1) We have zero ground stiffness, where the disks are free to self-assemble (i. e., R = 0). (2) We have the same disk, but with four fixed magnets at the corners of the unit cell resulting in positive stiffness ratio (i. e., R=+4). (3) In addition to the fixed four corner magnets, we add two fixed magnets to the center of the horizontal bars in the unit cell boundaries to achieve an overall negative stiffness ratio (i. e., R=-0.5) (Fig.\ref{fig:2D magnetic} (c) I). This design provides an anomaly on the longitudinal branch in $\Gamma-X$ segment. For corner magnets, we use $3 \times 2$ magnets with magnetic constants:$A=3.279 \times 10^{-11} N/m^\gamma$, $\gamma$ = -4. For center magnets, we use $3 \times 1$ magnets with magnetic constants: A=$8.6978 \times 10^{-12} N/m^\gamma$, $\gamma$ = -4.1973.

\end{document}